\documentclass[a4paper]{jpconf}
\usepackage{graphicx}
\usepackage{amsmath,amsfonts,amssymb,graphicx,cite}
\usepackage[paper=letterpaper,margin=1.2in]{geometry}
\begin{document}

%%%%%%%%%%%%%%%%%%%%%%%%%%%%%%%%%%%%%%%%%%%%%%
\newcommand{\im}{\mathop{{\rm Im}}}
\newcommand{\re}{\mathop{{\rm Re}}}
\newcommand{\cO}{{\cal O}}
\newcommand{\cS}{{\cal S}}
\newcommand{\cL}{{\cal L}}
\newcommand{\cT}{{\cal T}}
\newcommand{\cH}{{\cal H}}
\newcommand{\nn}{{\nonumber}}
\newcommand{\IR}{\mathbb{R}}
\newcommand{\IC}{\mathbb{C}}
\newcommand{\IZ}{\mathbb{Z}}
\newcommand{\id}{\mathbb{I}}
\newcommand{\comment}[1]{}

\newcommand{\be}{\begin{equation}}
\newcommand{\ee}{\end{equation}}
\newcommand{\bea}{\begin{eqnarray}}
\newcommand{\eea}{\end{eqnarray}}
\newcommand{\ba}{\begin{array}}
\newcommand{\ea}{\end{array}}
\newcommand{\ben}{\begin{enumerate}}
\newcommand{\een}{\end{enumerate}}
\newcommand{\ei}{\end{itemize}}
\newcommand{\bc}{\begin{center}}
\newcommand{\ec}{\end{center}}
\newcommand{\bt}{\begin{table}}
\newcommand{\et}{\end{table}}
\newcommand{\btab}{\begin{tabular}}
\newcommand{\etab}{\end{tabular}}
\newcommand{\todo}[1]{\noindent{\em #1}\marginpar{$\Longleftarrow$}}
\newcommand{\dd}{{\rm d}}

\newtheorem{theorem}{\bf THEOREM}
\def\thetheorem{\thesection.\arabic{theorem}}
\newtheorem{conjecture}{\bf CONJECTURE}
\def\thetheorem{\thesection.\arabic{conjecture}}
\newtheorem{proposition}{\bf PROPOSITION}
\def\thetheorem{\thesection.\arabic{proposition}}

\def\theequation{\thesection.\arabic{equation}}
\newcommand{\setall}{\setcounter{equation}{0}\setcounter{theorem}{0}\setcounter{table}{0}\setcounter{footnote}{0}}
\newcommand{\setequation}{\setcounter{equation}{0}}

%%%%%%%%%%%%%%%%%%%%%%%%%%%%%%%%%%%%%%%%%%%%%%%

\title{On the Physics of the Riemann Zeros}

\author{Yang-Hui He${}^{1}$, Vishnu Jejjala${}^{2}$ and Djordje Minic${}^{3}$}

\address{
${}^{1}$Rudolf Peierls Centre for Theoretical Physics, Oxford University, 1 Keble Road, OX1~3NP, U.K.,
Merton College, Oxford, OX1~4JD, U.K.,
Department of Mathematics, City University, Northampton Square, London EC1V~0HB, U.K.\ and
School of Physics, NanKai University, Tianjin, 300071, P.R.~China;
${}^{2}$Centre for Research in String Theory, Department of Physics, Queen Mary, University of London, Mile End Road, London E1~4NS, U.K.;
${}^{3}$IPNAS, Department of Physics, Virginia Tech, Blacksburg, VA 24061, U.S.A.}

\ead{${}^{1}$yang-hui.he@merton.ox.ac.uk,${}^{2}$v.jejjala@qmul.ac.uk,${}^{3}$dminic@vt.edu}

\begin{abstract}
We discuss a formal derivation of an integral expression for the Li coefficients associated with the Riemann $\xi$-function which, in particular, indicates that their positivity criterion is obeyed, whereby entailing the criticality of the non-trivial zeros. We conjecture the validity of this and related expressions without the need for the Riemann Hypothesis and
discuss a physical interpretation of this result within the Hilbert--P\'olya approach.
In this context we also outline a relation between string theory and
the Riemann Hypothesis.
\end{abstract}

\section{Introduction: Riemann Zeros and the Li Criterion}

This talk is devoted to the physics of the non-trivial zeros of the Riemann $\xi$-function,
the so-called Riemann zeros~\cite{ours}.
First, we review Li's Criterion, with an emphasis on requisite properties of the Riemann $\xi$-function.
In the central section of the talk, we show how using the exact formula for the
number of non-trivial zeros $N(T)$ up to a given height in the critical strip, and its derivative as eigenvalue distributions in the critical strip, would imply the positivity of Li's coefficients.
Finally, we interpret this result from a physicist's perspective emphasizing a
possibly exciting connection to string theory~\cite{future}.

%%%%%%%%%%%%%%%%%%+++++++++++++++++++++++++------------------------
%%%%%%%%%%%%%%%%%%+++++++++++++++++++++++++
%\section{Li's Positivity Criterion}\setall
Let us, perhaps more for the sake of notation, first remind the reader of the statement of the Riemann Hypothesis~\cite{riemann, e, t}:
The analytic continuation, from $z \in \IR_{>1}$ to the whole complex plane $z \in \IC$ of
$\zeta(z) := \sum\limits_{n=1}^\infty n^{-z}$
has all its zeros in the critical strip
$\cS := \{z \ | \ 0 < \re(z) < 1 \}$
lying on the critical line
$\cL := \{ z \ | \ \re(z) = \frac12 \}$.
Now, $\zeta(z)$ obeys a remarkable functional equation
$\zeta(z) = 2^z \pi^{z-1} \sin(\frac12 \pi z) \Gamma(1-z) \zeta(1-z)$,
which inspired Riemann to define the $\xi$-function, which will be central to our discussions:
\begin{equation}\label{xi}
\xi(z) := \frac12 z(z-1) \pi^{-\frac{z}{2}} \Gamma(\frac{z}{2}) \zeta(z) ~.
\end{equation}
There are many advantages to considering $\xi(z)$ instead of $\zeta(z)$, which we now summarize.

%%%%%%%%%%%%%------------------------------
\subsection{The Riemann $\xi$-function}
First, note that $\zeta(z)$ has trivial zeros at all negative even integers, which are conveniently cancelled by the corresponding simple poles of the $\Gamma$-function.
Hence, $\xi(z)$ has only non-trivial zeros, all located within the critical strip, and affords an elegant Weierstra\ss\ product expansion, known as the {\it Hadamard product}:
\begin{equation}\label{hadamard}
\xi(z) = \xi(0) \prod_{\rho \in \cS} (1 - \frac{z}{\rho}) ~,
\end{equation}
where $\xi(0) = \frac12$, and $\rho$ are the non-trivial zeros of $\zeta(z)$ in $\cS$ since none of the other factors in the definition of $\xi(z)$ vanishes therein.
It is an obvious but important fact that $\cS$ extends to both the upper and the lower half-planes and that all the $\rho$ occur in {\it conjugate pairs} above and below the real line.
We shall denote $\cS_+$ as the critical strip above the real line and correspondingly $\cL_+$ as the upper critical line.

Second, the functional equation becomes particularly simple:
\begin{equation}\label{fe}
\xi(z) = \xi(1-z) ~.
\end{equation}
Indeed, the symmetry about the critical line of $\re(z) = \frac12$ becomes manifest.
Hence, not only are the zeros symmetric about the real axis, they are also symmetric about the critical line.
The Riemann Hypothesis postulates that all the zeros in fact lie on $\cL$.
It is known that the first $10^{13}$ zeros lie on the critical line~\cite{gourdon}.

Third, consider the conformal mapping $z \mapsto \frac{z}{z-1}$, which maps $\cL_+$ to the boundary circle $|z|=1$ of the unit disk and the entirety of the left half of
$\cS_+$, {\em i.e.}, $\{z \ | \ 0 < \re(z) < \frac12 \}$,
to the interior of the open unit disk.
Consider, therefore, the function
\begin{equation}\label{phi-fe}
\phi(z) := \xi(\frac{z}{z-1}) = \xi(\frac{1}{1-z}) ~,
\end{equation}
where we have used \eqref{fe} in the second equality.
Since all zeros are symmetric around the critical line, to consider the zeros of $\xi$ within $\cS$ therefore amounts to considering the zeros of $\phi(z)$ within the open unit disk.
Indeed, the Riemann Hypothesis would require that there be no such zeros and indeed that all the critical zeros are on the boundary circle.
Since the number of zeros of an analytic function $f(z)$ in a region $R$ is equal to
$\int_R {\dd}z\ \frac{f'(z)}{f(z)}$,
we have that: The Riemann Hypothesis is equivalent to $\frac{\phi'(z)}{\phi(z)}$ being analytic ({\em i.e.}, holomorphic and without poles) within the open unit disk $|z| < 1$.

Fourth, a beautiful exact result relates the number of zeros up to height $T$ within $\cS$ and the $\xi$-function:
\begin{equation}\label{nt}
N(T) := \# \{ \rho \in \cS ~, 0 < \im(\rho) < T | \ \xi(\rho) = 0 \} =
\frac{1}{\pi} \im \log \xi (\frac12 + i T) ~.
\end{equation}
Extensive work has been done in estimating \eqref{nt}.
We have, using the definition \eqref{xi} for the first equality, that
\begin{equation}\label{eq:asym}
N(T) = \frac{1}{\pi} \vartheta(T) + 1 + \frac{1}{\pi} \im \log \zeta(\frac12 + i T) \sim \frac{T}{2\pi} \log \frac{T}{2\pi} - \frac{T}{2\pi} - \frac78 + \cO(\log T) ~,
\end{equation}
where historically $\vartheta(T) = \im \log \Gamma(\frac{i}{2} T + \frac14) - \frac{T}{2} \log \pi$ is known as the average part and $\frac{1}{\pi} \im \log \zeta(\frac12 + i T) \sim \cO(\log T)$ is the fluctuating part around the essentially $T \log T$ growth of the former.
This is the inspiration behind classical (the average) and quantum mechanical (the fluctuation) interpretations of the critical zeros~\cite{hp, berry}.

We note that $N(T)$ is a real step function, increasing by unity each time a new critical zero is encountered:
\begin{equation}\label{NT}
N(T) = \sum_{\rho \in \cS_+} \theta(T - \im \rho) = \sum_{j = 1}^\infty \alpha_j\ \theta(T - \mu_j) ~.
\end{equation}
It is vital to explain the above in detail.
We can explicitly write the upper critical zeros as $\rho_{j,k} = \sigma_{j,k} + i \mu_j$, indexed by $j, k\in \IZ_{>0}$, where $\sigma_{j,k}\in (0,1)$ and $\mu_j\in \IR_{>0}$.
The zeros are ordered so that $\mu_{j+1} > \mu_j$.
Crucially, as we do not assume the Riemann Hypothesis, the real part of $\rho$ need not be $\frac12$.
Furthermore, we do not assume simplicity of the zeros, which is as yet also unknown~\cite{conrey2}.
If, for example, we have a double root, we explicitly count this twice.
By the functional identity, zeros of the $\xi$-function not on the critical line are paired within the upper critical strip:
if $\sigma + i \mu$ is a zero, then so is $(1-\sigma) + i\mu$.
The index $k=1,\ldots,\alpha_j$ enumerates the zeros with the same imaginary part.
We order the zeros so that $\sigma_{j,k+1} \ge \sigma_{j,k}$.
The $\alpha_j$ then counts the number of zeros with imaginary part $\mu_j$ including the multiplicities of the roots.
If, for example, there are a pair of simple roots with imaginary part $\mu_j$, then in an epsilon interval around $\mu_j$, the counting function $N(T)$ jumps by two.
Contrariwise, if $\rho\in \cL_+$ is the only simple root with imaginary part $\mu_j$, then $N(T)$ jumps by one.
We know that the number of roots with imaginary part in the interval $(0,T)$ is finite.
Indeed, the asymptotics of the expression are given in \eref{eq:asym}.
In summary, $N(T)$ is a strictly increasing step function as we move up in height $T$ regardless of the Riemann Hypothesis.
The Riemann Hypothesis is, of course, the statement that $\sigma_j = \frac12$ for all zeros.

The rewriting of $N(T)$ lends itself to a wonderful interpretation:
$N(T)$ is a cumulative density function defined over the critical strip.
In other words, its derivative, $\widetilde{\rho}(T) := N'(T)$ is a density function of distributions.
That is, one could conceive of a physical system whose energy levels (eigenvalues of the Hamiltonian) are thus distributed;
this is along the school of thought of Hilbert--P\'olya~\cite{hp}.
Of course, the resulting eigenvalue density is highly non-smooth, but is, rather, a sum of delta-functions:
\begin{equation}\label{density}
\widetilde{\rho}(T) := N'(T) = \sum_{j = 1}^\infty \alpha_j\ \delta(T - \mu_j) ~.
\end{equation}

%%%%%%%%%%%%%%%================
\subsection{Li's Criterion}
Another striking property of the $\xi$-function was noted by Li~\cite{li} not too long ago.
Let $\{k_n\}$, for positive integers $n$, be defined by
\begin{equation}
k_n := \frac{1}{(n-1)!} \frac{\dd^n}{\dd z^n} \left[ z^{n-1} \log \xi(z) \right]_{z=1} ~;
\end{equation}
then we have that:
The Riemann Hypothesis is equivalent to the condition that $k_n \ge 0$ for all $n \in \IZ_{>0}$.
This is Li's Criterion.
Li also showed two equivalent ways of writing these numbers, namely
\begin{eqnarray}
\label{kn-phi}
\frac{\phi'(z)}{\phi(z)} &=& \sum_{n=0}^\infty k_{n+1} z^n ~; \\
\label{kn-rho}
k_n &=& \sum_{\rho \in \cS} \left[ 1 - \left(1 - \frac{1}{\rho}\right)^n \right] ~,
\end{eqnarray}
where we recall that
$
\phi(z) := \xi(\frac{1}{1-z}) $.

\section{Density Function for the Distribution of Critical Zeros}
\setequation

Thus armed with all necessary ingredients, let us proceed to rewrite the Li coefficients in a suggestive form. We will do so {\it formally} and point out the subtleties involved at the end of the manipulations~\cite{ours}.

First, integrating \eqref{kn-phi} and using the definition as well as the functional equation \eqref{phi-fe}, we have that, for $z$ in the unit disk,
\begin{equation}\label{xi-expansion}
\log \xi(\frac{z}{z-1}) =
\log \xi(\frac{1}{1-z}) =
\sum_{n=1}^\infty k_{n} \int {\dd}z\ z^{n-1} =
-\log 2 + \sum_{n=1}^\infty \frac{k_{n}}{n} z^n ~,
\end{equation}
where $-\log 2$ is easily checked to be the constant of integration.

We note, however, that the above expansion has radius of convergence strictly less than unity, whereby excluding the unit circle, to which, crucially, $\cL$ is mapped under our conformal transformation $z \mapsto \frac{z}{z-1}$.
Therefore, it is imperative to analytically continue.
Let us start with \eqref{xi-expansion} and rewrite the expansion about the point $z=-1$. Note that the region of convergence for this is an open of circle of radius two centered about $z=-1$, which, in particular, encloses the entirety of the closed unit disk, except the point $z=1$, where there is a pole for $\phi(z) = \xi(\frac{1}{1-z})$.
\comment{
This is important since under our conformal map $z \mapsto \frac{z}{z-1}$, the critical line is mapped to the boundary unit circle centered at the original, on which \eqref{xi-expansion} fails to converge.}
Therefore, we have that
\begin{equation}\label{xi-exp2}
\log \xi(\frac{1}{1-z}) =
\log \frac12 + \sum_{n=1}^\infty \frac{k_{n}}{n} z^n =
b_0 + \sum_{n=1}^\infty b_n (z+1)^n ~,
\end{equation}
where we have written the two expansions together for comparison.

Expanding $(z+1)^n$, we readily obtain an expression for the Li coefficients in terms of the new expansion coefficients $b_n$ (studied further in reference~\cite{coffey}),
$
k_n = n \sum_{j=n}^\infty {j \choose n} b_j ~.
$
\comment{
Therefore, showing that all $b_n \ge 0$ suffices to imply that all $k_n \ge 0$.
}
We now wish to solve for the $k_n$ explicitly and demonstrate positivity.

We first see, using the counting formula \eqref{nt} and the expansion \eqref{xi-exp2}, that
\begin{equation}\label{n-mu}
N(\mu) = \frac{1}{\pi} \im \log \xi(\frac12 + i \mu) =
\sum_{n=1}^\infty \frac{b_n}{\pi} \im \left(\frac{2\mu+i}{2\mu-i} + 1\right)^n ~,
\end{equation}
where we have used the substitution $\frac12 + i \mu = \frac{1}{1-z}$, or $z = \frac{2\mu+i}{2\mu-i}$.
Then, since
\begin{equation}
\im \left(\frac{2\mu+i}{2\mu-i} + 1\right)^n =
\frac{(4\mu)^n}{(4\mu^2+1)^{\frac{n}{2}}} \sin (n \tan^{-1} \frac{1}{2\mu}) =
2^n \cos^n\theta \ \sin(n \theta) ~, \qquad
\cos \theta := \frac{2\mu}{\sqrt{4\mu^2+1}} ~,
\end{equation}
and due to the definition of the Chebyshev polynomial of the second kind
\begin{equation}\label{chebyU}
U_{m-1}(\cos \phi) := \frac{\sin(m \phi)}{\sin \phi} ~,
\end{equation}
we arrive at the conclusion
that for all $m\in \IZ_{>0}$:
\begin{equation}\label{kn-int}
k_m = 32 m \int_0^\infty {\dd}\mu\ \frac{\mu}{(4\mu^2+1)^2}\ N(\mu)\ U_{m-1}(\frac{4\mu^2-1}{4\mu^2+1}) ~.
\end{equation}
This is a beautiful and remarkable integral expression for the Li coefficients
which works wonderfully numerically~\cite{ours}.
\comment{
Crucially, note that we have not invoked the Riemann Hypothesis in the above steps, but only the cumulative density function $N(\mu)$ for $\mu$ measuring the height within $\cS_+$ above the real axis.
}

Let us now proceed to simplify the integral \eqref{kn-int}.
First, recall that Chebyshev polynomials of the second kind are related to the Chebyshev polynomials of the first kind by the equation
\begin{equation}\label{int-TU}
\int {\dd}x \ U_n(x) = \frac{1}{n+1} T_{n+1}(x) ~.
\end{equation}
Then using the integral expression for $k_m$ we have that~\cite{ours}
\begin{equation}\label{kn-sum}
k_m = 2 \sum_{j=1}^\infty \alpha_j\ (1 - T_m(\frac{4\mu_j^2-1}{4\mu_j^2+1})) ~,
\quad m \in \IZ_{>0} ~.
\end{equation}

Recalling the definition of the Chebyshev polynomial of the first kind, that
$T_m(\cos \theta) := \cos (m \theta)$ and remembering that all $\alpha_j$, regardless of actual value, are strictly positive integers, each summand in \eqref{kn-sum} is non-negative.
In fact, because $\frac{4\mu_j^2-1}{4\mu_j^2+1} < 1$ strictly for any imaginary part as we know the first $\mu_j$ is at least $14$,
\begin{equation}
T_m(\frac{4\mu_j^2-1}{4\mu_j^2+1}) < 1 \quad
\mbox{ for any given } m \in \IZ_{>0}
\mbox{ and for all } j \in \IZ_{>0} ~,
\end{equation}
so each summand is actually strictly positive.
In conclusion, this would imply that all Li's coefficients $k_m > 0$.
Indeed, we have numerically evaluated~\cite{ours} the first few $k_m$ coefficients given in \eqref{kn-sum}, using the first $10^4$ critical zeros, and see that they are very close to the results obtained from \eqref{kn-phi} and \eqref{kn-int}, and moreover constitute an increasing sequence of strictly positive numbers.
In particular, the above discussions would force all critical zeros to lie entirely on $\cL$.

\comment{
Thus, if $(1 - T_m(\frac{4\mu_j^2-1}{4\mu_j^2+1}))$ is non-negative for all $\mu_j$, then a lower bound on $k_m$ is obtained by setting $\alpha_j = 1$:
\begin{equation}\label{kn-sum2}
k_m \ge 2 \sum_{j=1}^\infty (1 - T_m(\frac{4\mu_j^2-1}{4\mu_j^2+1})) ~,
\quad m \in \IZ_{>0} ~.
\end{equation}
Crucially, we have arrived at the conclusion that $k_m$ are real and non-negative {\it without} (unlike in the necessary direction of the proof in Li's Criterion) assuming the Riemann Hypothesis.
%Li's Criterion therefore implies that the lower bound on $k_m$ is exact, and $\alpha_j = 1$ for all $j$.
Therefore, our arguments seem indeed to compel all critical zeros of the Riemann zeta function to lie upon the critical line.
}

We seem to have arrived the positivity of the Li's coefficients without the assumption of the Riemann Hypothesis! This, of course, is not quite true and the criticality of the zeros of the $\xi$-function had in fact been inherently invoked in the choice of a branch cut for the logarithm. Nevertheless, we have {\it formally} arrived the integral formula \eqref{kn-int} which does not seem to depend on the precise location of the zeros and the subsequent choice of branch-cut.
Note that Li's criterion can be applied to other zeta functions~\cite{li} and thus it is natural
to conjecture that our integral and summation formulas for the Li coefficients should exist in those cases as well.

\section{The Physics of the Critical Zeros and String Theory}
\setequation

Attempts to use a physical argument to explain the Riemann Hypothesis date back to Hilbert and P\'olya~\cite{hp, berry}.
Were one to find a matrix
\begin{equation}\label{H}
H = \frac12 \cdot \id + i\cdot \cT ~,
\end{equation}
whose eigenvalues are all the zeros of $\xi(z)$ and if in addition the matrix ${\cal T}$ were Hermitian, then all the zeros would lie on the critical line $\cL$.
Therefore, one would wish to find the appropriate operator $H$ and its corresponding eigenfunctions, or, in physical terms, the Hamiltonian and its wave functions.

What can one say about the quantum mechanical picture~\cite{ours}?
Let us investigate the secular equation of the operator in \eqref{H}.
It should by construction have roots corresponding to the zeros of $\xi(z)$, that is,
\begin{equation}\label{eq:wkb}
\xi(z) = \det (z \cdot \id - H) ~.
\end{equation}
To arrive at the appropriate matrix description, we start by considering the expansion \eqref{xi-expansion}, which, combined with \eqref{kn-sum}, gives
\begin{equation}
\log\xi(\frac{1}{1-s}) = -\log 2+\sum_{n=1}^\infty \sum_{j=1}^\infty \frac{2\alpha_j}{n} \left[ 1 - T_n( \frac{4\mu_j^2-1}{4\mu_j^2+1} ) \right] s^n ~.
\end{equation}
Put $\cos\theta_j := \frac{4\mu_j^2-1}{4\mu_j^2+1}$.
Using the definition of the Chebyshev polynomial and reversing the sums, we obtain
\be
\log\xi(\frac{1}{1-s})
= -\log 2 + \sum_{j=1}^\infty 2 \alpha_j \sum_{n=1}^\infty \left( \frac{1}{n} - \frac{1}{n}\cos(n \theta_j) \right) s^n.
%\nn
%&=& -\log 2+\sum_{j=1}^\infty \alpha_j \left[ -2\log(1-s) + \log(1-s \ e^{i \theta_j}) + \log(1-s\ e^{-i \theta_j}) \right] ~.
\ee
Substituting $z = \frac{1}{1-s}$ and using that
$e^{\pm i \theta_j} = 1 - (\frac12 \pm i \mu_j)^{-1}$,
we find
\be\label{eq:xi1}
\xi(z) = \frac{1}{2} \prod_{j=1}^\infty \left( \frac{(z-(\frac12 + i\mu_j))(z-(\frac12 - i\mu_j))}{(\frac12 + i\mu_j)(\frac12 - i\mu_j)} \right)^{\alpha_j} ~,
\ee
which is simply a refined version of the Hadamard product, a reassuring check.
Note that each factor in the product obeys the functional identity.

Equation~\eref{eq:xi1} is to be compared with \eref{eq:wkb}.
We see that there exists a matrix such that its eigenvalues coincide with $\mu_j$, which is consistent with the Hilbert--P\'olya picture.
As the expression vanishes for zeros on the critical line, the refined Hadamard product is already in the form of the determinant of an infinite matrix, and the eigenvalues of $\cT$ are real.

\subsection{Topological String Theory, Gromov--Witten Invariants and the Critical Zeros}

The profound proofs of the Riemann Hypothesis for the case of finite fields by
Weil, Deligne and others are essentially geometric, and thus one might expect a deeply geometric 
resolution of the Riemann Hypothesis~\cite{bombieri}.
Motivated by this expectation in this concluding section we present a possibly exciting connection between
Gromov--Witten (GW) invariants, topological string theory and the Riemann zeros~\cite{future}.

First we recall a reformulation of the Riemann Hypothesis in terms of
an elementary problem due to Lagarias~\cite{lagarias}:
Let $H_n = \sum_{j=1}^{n} \frac{1}{j}$ be the $n$-the harmonic number 
and let $\sigma_1 = \sum_{d|n} d$ is the sum of divisors (Ramanujan) function, then
the Riemann Hypothesis is true if and only if
\begin{equation}
\sigma_1(n) < H_n + e^{H_n} \log H_n
\end{equation}
for $n >1$.
In the following we notice the appearance of $\sigma_1(n)$ in the genus-one free energy
of a particular topological string theory~\cite{future}.

We start with the following statement from reference~\cite{mina}:
Take the type II string on the local Calabi--Yau threefold defined by 
\begin{equation}
uv = W(y,z) = y^2 - P_n(z)^2 + 1 \ ,
\end{equation}
where $W(y,z) = 0$ defines a  Riemann surface and $P_n(z)$ is a degree $n$ polynomial. The genus-one amplitude can be given exactly (due to Seiberg--Witten (SW) theory)  by
\begin{equation}
F_1(\tau) = - \log \eta(\tau) \ ;
\end{equation}
this is also the genus-one free energy of the topological string (i.e. topologically twisted from the ${\cal N} = 2$ SW theory). From the definition of the Dedekind eta-function, we can expand, setting $q = \exp(2\pi i \tau)$,
\begin{eqnarray}
\nn 
F_1(\tau) &=& - \log \eta(\tau) = -\frac{1}{24} \log q 
- \sum\limits_{i=1}^\infty \log(1 - q^i) 
= -\frac{\pi i \tau}{12} - \sum\limits_{i=1}^\infty\sum_{j=1}^\infty \frac{q^{ij}}{j}\\
&=& -\frac{\pi i \tau}{12} - \sum\limits_{i=1}^\infty\sum_{j=1}^\infty \frac{q^{ij}}{j}
= -\frac{\pi i \tau}{12} - \sum\limits_{n=1}^\infty(\sum_{d|n} d )q^{n}
= -\frac{\pi i \tau}{12} - \sum\limits_{n=1}^\infty \sigma_1(n) q^{n} \ .
\end{eqnarray}
Thus, the GW invariants at genus $g=1$ and number of marked points $m=0$, for the above Calabi--Yau $X$ is the Ramanujan $\sigma_1$
function! That is:
\begin{equation}
N_{g=1,m=0}(n; X) = \sigma_1(n) \ .
\end{equation}
Then, the statement, due to Lagarias' theorem~\cite{lagarias}, is that the asymptotics of the free energy of this particular topological string at genus-one world-sheet 
embedding should be $\sigma_1(n) < H_n + e^{H_n} \log H_n$, if and only if Riemann Hypothesis holds. 
Therefore: (1), we know that the Lagarias bound on $\sigma_1(n)$ in terms of the harmonic numbers is equivalent to Riemann Hypothesis and (2), we know of examples of GW invariants which are equal to $\sigma_1(n)$. Hence the question: does there exist an independent geometric bound on such invariants?
(At this point one might wonder whether the GW invariants could encode the critical zeros. This seems unnatural because GW invariants are integers indexed by integers whilst the critical zeros are not, whence it is hard to imagine, though no entirely impossible, that a combinatorial problem should be associated thereto.)

Also, by considering rational curves of degree $d$ on the K3 surface we can deduce that~\cite{future}
$
\sum\limits_{n=0}^k \tau(n) N_1(k-n) \le k \sigma_1(n),
$
for each $k = 0,1,2,\ldots,$
and thus restate the Riemann Hypothesis in the following form~\cite{future}:
\be
\sum\limits_{n=0}^k \tau(n) N_1(k-n) \le k(H_k + e^{H_k} \log H_k), \quad k=0,1,2,\ldots
\ee
where $\tau(n)$ is the famous Ramanujan multiplicative tau-function (for which, $\tau(0) = 0$), and $N_1(j)$ is the number of elliptic curves with $j$ nodes and passing through a given point on a generic, smooth K3 surface.
Note that according to reference~\cite{oxford}
there exists a universal constant $a$ such that
\begin{equation}
Rat(X) \le \chi(X) + a \ , 
\end{equation}
for every generic smooth surface $X$, with $\chi(X)$ being the Euler number.
We need a version of this statement for elliptic curves.

%Upon this issue of an independent geometrical bound there appears to exist some literature of %potential pertinence.
The papers on the so-called Calabi--Yau differential equations have studied the integer coefficients of mirror maps, which we know to be an interesting phenomenon in and of itself. 
(In particular, the mirror symmetry might be related to the
Birch and Swinnerton-Dyer conjecture~\cite{yang}.) Reference~\cite{mirror} is particularly interesting.
We know that, after judicious transformation, the coefficients of the mirror map are precisely the GW invariants.
In reference~\cite{mirror}, an expression of the coefficients in terms of harmonic numbers is given.
The strategy is clear.
There appear to exist two directions: one giving the GW invariants in terms of $\sigma_1(n)$ and the other, the GW invariants via the mirror map, in terms of the harmonic numbers. One thus must relate the two and the question is whether the relation is strong enough to give the Lagarias bound.

\ack{
We are indebted to Michael Berry, Hung Bui, Philip Candelas, Mark Coffey, Roger Heath-Brown, Jon Keating, Eric Sharpe, Al Shapere and
Andrew Wiles for interesting conversations and questions regarding the material presented in this talk. YHH is supported in part by an Advanced Fellowship from the STFC, U.K.\ in association with the Rudolf Peierls Centre for Theoretical Physics, University of Oxford, a Supernumerary Fellowship of Merton
College, Oxford, and impending funding from City University, London and a Chang-Jiang
Chair Professorship from the Chinese Ministry of Education, at Nan-Kai University, Tian-Jin.
VJ is supported by STFC.
DM is supported in part by the U.S.\ Department of Energy under contract DE-FG05-92ER40677.
DM thanks the organizers of the QTS6 conference for the opportunity to present this work.
VJ and DM also wish to thank the magnanimity of the Warden, Fellows, and Scholars of Merton College, Oxford University.
}

\end{document}